\documentclass[
    aps,
    pra,
    reprint,
    superscriptaddress
]{revtex4-2}

\usepackage[english]{babel}

\usepackage{amsmath}
\usepackage{amsfonts}
\usepackage{amssymb}
\usepackage{bm}
\usepackage{bigints}

\usepackage{graphicx}
\usepackage{dcolumn}

\usepackage{siunitx}

\usepackage{xcolor}

\usepackage{hyperref}

\AtBeginDocument{\RenewCommandCopy\qty\SI}

\begin{document}

\preprint{APS/123-QED}

\title{Dynamics of repeated BEC formation and extraction in dimple traps}

\author{Kyrylo Kovalchuk}
\affiliation{Department of Physics, Taras Shevchenko National University of Kyiv, 64/13, Volodymyrska Street, Kyiv 01601, Ukraine}

\author{Dominik Pfeiffer}
\affiliation{Technische Universit\"at Darmstadt, Institut f\"ur Angewandte Physik, Schlossgartenstraße 7, 64289 Darmstadt, Germany}

\author{Ludwig Lind}
\affiliation{Technische Universit\"at Darmstadt, Institut f\"ur Angewandte Physik, Schlossgartenstraße 7, 64289 Darmstadt, Germany}

\author{Mark Edwards}
\affiliation{Department of Biochemistry, Chemistry, and Physics, Georgia Southern University, Statesboro, Georgia 30460-8031, USA}

\author{Alexander Yakimenko}
\affiliation{Department of Physics, Taras Shevchenko National University of Kyiv, 64/13, Volodymyrska Street, Kyiv 01601, Ukraine}
\affiliation{Dipartimento di Fisica e Astronomia Galileo Galilei, Universit\'a di Padova, and INFN, Sezione di Padova,
Via Marzolo 8, 35131 Padova, Italy}

\author{Gerhard Birkl}
\affiliation{Technische Universit\"at Darmstadt, Institut f\"ur Angewandte Physik, Schlossgartenstraße 7, 64289 Darmstadt, Germany}
\affiliation{Helmholtz Forschungsakademie Hessen f\"ur FAIR (HFHF), GSI Helmholtzzentrum für Schwerionenforschung, 64291 Darmstadt}

\date{\today}

\begin{abstract}
{
We investigate repeated Bose–Einstein-condensate (BEC) formation and extraction in a dimple trap embedded in a reservoir of thermal atoms using a kinetic model. The model includes pulsed extraction, evaporation, three-body losses, and thermal-atom replenishment. Three extraction protocols are compared: extraction of all atoms from the dimple (BEC and thermal atoms), full and partial extractions of the BEC, but not of the thermal atoms. Residual atoms in the dimple after extraction seed subsequent Bose-stimulated growth and reduce the recovery time between extractions, but also enhance density-dependent losses. For all protocols, repeated extraction of BECs can be achieved without replenishment, but the number of BEC formations is limited by reservoir depletion and heating. With continuous replenishment, the system can reach a periodic steady-state regime, after an initial transient period, controlled by the externally imposed rates of extraction pulses and thermal-atom input. Within the explored parameter range, partial BEC extraction gives the highest efficiency, particularly for short extraction periods and high input rates. These results identify seeding by residual populations of BECs and thermal atoms as a kinetic mechanism for improving repeated condensate production in dimple traps.
}

\end{abstract}

\maketitle

\section{Introduction}
Since the first realization of Bose--Einstein condensation (BEC) in dilute atomic gases~\cite{Anderson1995Observation,PhysRevLett.75.3969,PhysRevLett.75.1687}, ultracold quantum matter has provided a highly controllable platform for studying nonequilibrium condensate formation and finite-temperature dynamics. Kinetic and finite-temperature theories have clarified how elastic collisions, Bose stimulation, rethermalization, evaporation, and loss processes govern condensate growth in trapped gases~\cite{luiten1996kinetic,pethick_smith_2008,griffin_nikuni_zaremba_2009,RevModPhys.71.463,PhysRevA.62.063608,Gardiner_1997,Bijlsma2000}. The same mechanisms constrain repeated or continuously replenished condensate production, where heating, depletion, and finite thermalization rates limit the achievable condensate flux and duty cycle.


A system where small BECs are created and made available to other systems on a regular, periodic basis would have several advantages. Quantum sensors (e.g., using atom interferometry) that involve coherent matter waves sourced by a BEC for their operation use destructive imaging for measurement and need newly formed condensates for the next measurement. Access to a regular, periodic source of BECs could also stimulate the use of condensates in other applications such as atomic clocks and atom lasers. Repeated sample production with low deadtime approaching continuous operation might even be mandatory for some applications, e.g., navigation. Phase distortions caused by mean-field interaction might even favor small condensates at high repetition rates over large condensates with low rates with respect to signal-to-noise considerations. Thus, any application involving repeated use of Bose-Einstein condensates would benefit from the availability of such a system \cite{meng:2024:closed,geiger:2020:highaccuracy}.

The need for stable, replenishable condensate sources has driven efforts to develop coherent matter-wave sources, including atom lasers based on controlled outcoupling from a trapped condensate~\cite{Hagley1999AtomLaser,Helmerson_1999,bolpasi:2014:atomlaser},  as well as related schemes for coherent atom injection and extraction in tailored trapping geometries~\cite{Loiko2014}. Pulsed and quasi-continuous atom-laser experiments \cite{ROBINS2013265,Chikkatur2002Continuous}, together with related work on guided atomic beams and evaporative cooling~\cite{PhysRevA.72.033411}, have demonstrated the feasibility of coherent atom extraction while also exposing the kinetic constraints on high-duty-cycle operation. In particular, sustained operation requires condensate replenishment to compensate for outcoupling and losses without introducing excessive heating or reservoir depletion. Finite reservoir size, density-dependent losses, and limited rethermalization rates therefore restrict the achievable flux and repetition rate.

Related approaches based on continuous loading and Bose-stimulated growth \cite{Chen2022,Chikkatur2002Continuous,PhysRevA.57.2030} provide access to nonequilibrium steady states, but are likewise limited by dissipation, spatial inhomogeneity, and incomplete rethermalization. In this context, spatially structured potentials such as dimple traps and hybrid reservoir-core configurations have been explored as a means to enhance condensation efficiency and local phase-space density \cite{PhysRevLett.110.263003,PhysRevA.83.013630,PhysRevA.91.013601}. \textcolor{black}{Experimental studies of optimized BEC production in dipole traps have also demonstrated the importance of loss processes for condensate production~\cite{PhysRevA.84.043641}.} These systems allow for controlled particle flux into a tightly confined region, accelerating condensate growth; however, even in such optimized geometries, the overall efficiency remains limited by the interplay between growth kinetics, depletion, and transport processes.

An alternative and experimentally accessible strategy is based on repeated extraction of the condensate from a partially replenished system. In such protocols, a condensate is formed, a fraction of it is extracted, and the remaining atomic cloud serves as a reservoir for subsequent recondensation. While conceptually related to continuously pumped and reservoir-fed configurations \cite{PhysRevA.68.023607,Chikkatur2002Continuous}, these approaches differ in that the system operates in a cyclic rather than steady-state regime. Existing realizations of continuous or quasi-continuous condensation rely on maintaining a balance between gain and loss in a near-steady configuration \cite{Chen2022,CHEN2023361}, which typically requires stringent conditions on cooling, loading, and isolation. In contrast, cyclic operation relaxes these constraints but introduces a fundamentally different dynamical regime, where condensate growth, depletion, and rethermalization are strongly coupled over successive cycles.

Despite conceptual similarities to replenished and continuously pumped systems, the regime of repeated condensate extraction has received comparatively little direct attention in the literature. Previous studies have primarily focused on achieving steady-state gain and coherence \cite{Bhongale2000}, or on enhancing phase-space density and condensate growth in localized regions such as dimple traps \cite{PhysRevLett.81.2194,PhysRevLett.110.263003,PhysRevA.73.053603}, rather than on the dynamics of repeated partial depletion and regrowth. \textcolor{black}{Theoretical studies based on quantum kinetic models have addressed continuous atom-laser operation and the associated condensate growth and loss dynamics~\cite{PhysRevA.86.013640}.} As a result, it remains unclear whether the presence of a residual condensate fraction can significantly accelerate subsequent growth via Bose-enhanced processes, or whether cumulative heating and depletion effects ultimately limit performance. \textcolor{black}{In this cyclic regime, the residual dimple population seeds subsequent Bose-stimulated growth, but also enhances density-dependent losses and modifies the reservoir temperature.}

This motivates a kinetic analysis of repeated extraction protocols in which the condensate is not necessarily removed completely after each cycle. The central questions are how partial depletion modifies subsequent condensate growth, whether residual dimple populations can act as a \textcolor{black}{seed} that accelerates Bose-stimulated regrowth, and how this benefit competes with cumulative heating, depletion, and density-dependent losses. It is also important to determine under which conditions cyclic operation can improve the cumulative condensate output relative to single-shot extraction and approach a reproducible high-duty-cycle regime. In this work, we address these questions within a reservoir-dimple population-kinetic model by comparing several extraction protocols with and without external reservoir replenishment. This enables us to identify the kinetic operating regimes in which repeated extraction can provide an efficient route toward high-duty-cycle condensate production, relevant to pulsed or continuous coherent matter-wave sources.

The paper is organized as follows. Section~\ref{sec:model} describes the kinetic model and the extraction/replenishment protocols; Sec.~\ref{bec_extr} presents the results for repeated extraction with and without external replenishment; and Sec.~\ref{sec:conlc} summarizes this work and draws relevant conclusions.

\section{Model}\label{sec:model}
\subsection{Reservoir-dimple kinetic description}
\begin{figure}[htb]
     \centering \includegraphics[width=\columnwidth]{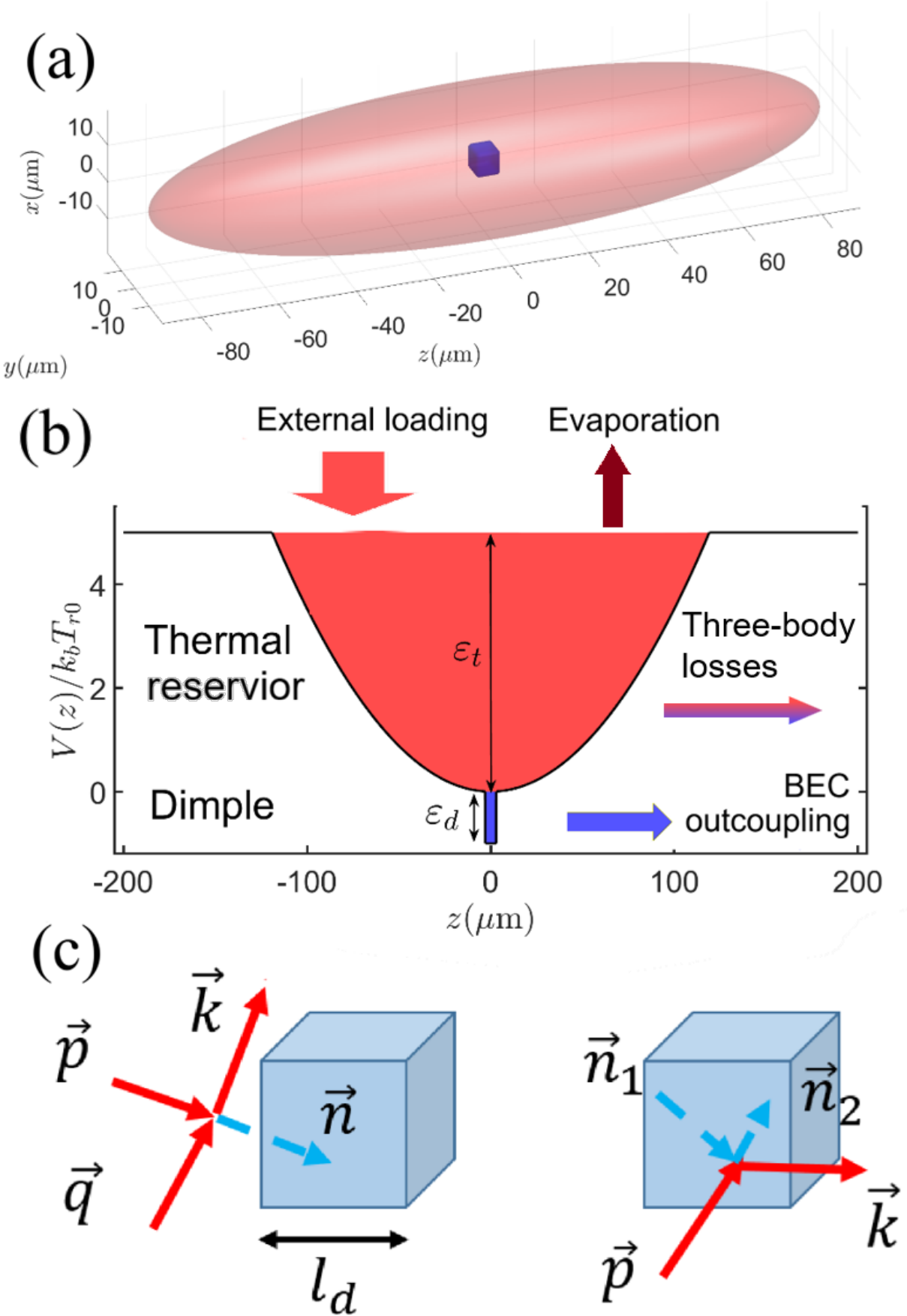}
\caption{
(a) Schematic of the trap geometry: a localized dimple potential (blue) is embedded in a harmonic thermal reservoir (red).  The reservoir density is represented by an ellipsoidal isodensity surface for initial particle number \(N_{r0}=3\times10^{5}\) and temperature \(T_{r0}=150\,{\rm nK}\). (b) Axial cross section of the trapping potential. Reservoir states occupy the energy range \(0<E<\varepsilon_t\), whereas dimple states lie in the interval \(-\varepsilon_d<E<0\). (c) Collision processes included in the kinetic model: solid red arrows denote reservoir atoms and dashed blue arrows denote dimple atoms.
}\label{sch}
\end{figure}

We use the reservoir-dimple kinetic model developed in Ref.~\cite{PhysRevA.91.013601} as the starting point for the present analysis. Since the extraction and replenishment protocols introduced below act directly on the reservoir and dimple populations, we briefly recall the main elements of the model and establish the notation used throughout this work. 

The system consists of a large thermal reservoir coupled to a tightly confined dimple trap, as illustrated in Fig. \ref{sch}. The reservoir is modeled as an isotropic harmonic trap with frequency \(\omega\), and the thermal cloud is described by a Boltzmann distribution truncated at the trap depth \(\varepsilon_t\), with initial temperature \(T_{r0}\equiv T_r(t=0)\). It is important to note that the frequency \(\omega\) used in the reservoir model represents the geometric mean of the harmonic trapping frequencies. Therefore, the model is not restricted to a physically isotropic reservoir, but applies to anisotropic harmonic traps satisfying
\begin{equation}
    \omega_x\omega_y\omega_z = \omega^3.
    \label{cond1}
\end{equation}

The dimple is treated as a localized attractive well of effective volume \(l_d^3\) \textcolor{black}{and depth} \(\varepsilon_d\). Following the kinetic formulation of Ref.~\cite{PhysRevA.91.013601}, we use an auxiliary box-like representation of the dimple spectrum to define the density of states entering the population kinetics. This representation is not intended as a detailed model of the optical dimple potential. As discussed in Ref.~\cite{PhysRevA.91.013601}, alternative choices for the dimple spectrum, such as a harmonic oscillator spectrum, lead to similar dynamical behavior. In the long-time dynamics with repeated extractions considered here, the qualitative dynamics are therefore expected to be governed mainly by the density of states and by collision-induced redistribution among dimple levels, rather than by the precise form of the single-particle eigenfunctions.

A central assumption of the kinetic model is that the thermal cloud remains close to equilibrium and is spatially much larger than the dimple. Under these conditions, the dimple dynamics are determined primarily by the local thermal density and temperature near the trap center, together with the reservoir density of states. Consequently, the detailed global structure of the reservoir is unimportant, provided that the thermal-cloud dimensions remain much larger than the dimple size in all directions.

To assess the validity of this condition quantitatively, we characterize the reservoir size using the effective thermal volume defined in Ref.~\cite{PhysRevA.91.013601}:
\begin{equation}
    V_{r0} = \left( \frac{2\pi k_B T_{r0}}{ m \omega^2} \right)^{3/2}\frac{\sqrt{\pi}}{4}\,\frac{\gamma\!\left(3,\varepsilon_t/k_B T_{r0}\right)}{\gamma\!\left(3/2,\varepsilon_t/k_B T_{r0}\right)} ,
\end{equation}
where \(m\) is the atomic mass and \(\gamma(\nu,x)\) denotes the lower incomplete gamma function. For \(\nu \gtrsim 5\), the latter may be approximated by the complete gamma function, yielding:
\begin{equation}
    V_{r0} \approx \left( \frac{2\pi k_B T_{r0}}{ m \omega^2} \right)^{3/2} = L_x L_y L_z,
\end{equation}
\begin{equation}
    L_i = \left( \frac{2\pi k_B T_{r0}}{m \omega_i^2} \right)^{1/2}, \quad \min \left( L_i \right) \gtrsim l_d,
    \label{cond2}
\end{equation}
where \(i\) labels the Cartesian axes. Under these conditions, the simulation results are expected to be the same for different sets of trap frequencies, provided that conditions (\ref{cond1}) and (\ref{cond2}) are satisfied.

\begin{table}[t]
    \centering
    \caption{Parametric set used for the simulations}
    \label{tab}
    \begin{tabular}{lll}
        \hline
        Symbol & Parameter & Value \\
        \hline
        \(\omega_{x,y}\) & Radial trap frequency & \(2\pi\times70.5~\mathrm{Hz}\) \\
        \(\omega_z\) & Axial trap frequency & \(2\pi\times16.0~\mathrm{Hz}\) \\
        \(\omega\) & Trap frequency (geom. mean) & \(2\pi\times43.0~\mathrm{Hz}\) \\
        \(l_d\) & Dimple size & \(5~\mu\mathrm{m}\) \\
        \(\varepsilon_d\) & Dimple depth & \(k_B\times150~\mathrm{nK}\) \\
        \(N_{r0}\) & Initial reservoir population & \(3\times10^5\) \\
        \(T_{r0}\) & Initial reservoir temperature & \(150~\mathrm{nK}\) \\
        \hline
    \end{tabular}
\end{table}

\textcolor{black}{The parameters and initial conditions used throughout the simulations are summarized in Table ~\ref{tab}. Unless explicitly stated otherwise, these parameters are kept constant, while the reservoir trap depth \(\varepsilon_t\) is varied only for the simulations in which its dependence is considered. This allows the effects of atom extraction from the dimple and external replenishment of the reservoir to be considered independently of changes in the underlying initial state.}

\textcolor{black}{The resulting trap configuration is illustrated in Fig.~\ref{sch}(a). The corresponding cross section of the trapping potential is shown in Fig.~\ref{sch}(b). For the initial reservoir temperature specified in Table \ref{tab}, the characteristic dimensions of the thermal cloud are \(L_x=L_y=21.4~\mu{\rm m}\) and \(L_z=94.3~\mu{\rm m}\). The effective-volume ratio between the thermal reservoir and the dimple is \(V_{r0}/l_d^3\approx350\) in the limit \(\varepsilon_t\rightarrow\infty\) and \(V_{r0}/l_d^3\approx310\) for \(\varepsilon_t=k_B\times750~{\rm nK}\), demonstrating that the reservoir is substantially larger than the dimple for the parameters considered.}

To model the non-equilibrium dynamics, two different types of two-body elastic collisions are considered. In the first type, a collision between two atoms in the reservoir transfers one of the atoms to the dimple, while the other atom gains energy. A schematic representation of such collisions is shown in Fig.~\ref{sch}(c), left panel. As a result of this interaction the second atom might also leave the system if its energy exceeds the depth of the reservoir, i.e., \(E>\varepsilon_t\). In this case the collision becomes irreversible. If this does not happen, then the reversible case, where the result of the collision is that both atoms remain in the reservoir, is also included.

In the second type of collision, a thermal atom collides with an atom from the dimple, changing the energies of both atoms. A schematic representation of one such process is shown in Fig.~\ref{sch}(c) (right panel), where the dimple atom remains trapped but is transferred to a different dimple state. Depending on the energy exchanged during the collision, the dimple atom may also acquire sufficient energy to leave the dimple. The corresponding reverse processes are included as well. Whereas the first type of collision increases the population of atoms inside the dimple, the second type primarily drives thermalization by redistributing the population among dimple energy levels and, when energetically allowed, by coupling dimple states to the thermal reservoir.

Since these collisions occur predominantly at the interface between the dimple and the reservoir, i.e., near \({\bf r}=0\), the corresponding rate equations can be integrated over momentum, creating a set of energy-dependent kinetic equations for the condensate fraction \(f_0\), the non-condensate dimple distribution function \(f(\tilde{E},t)\), and the thermal number and energy fractions \(f_r=N_r(t)/N_{r0}\) and \(e_r=E_r(t)/E_{r0}\), where \(E_{r0}\equiv E_r(t=0)\)~\cite{PhysRevA.91.013601}. The dimple-state energies in the range \(-\varepsilon_d<E<0\) are then rescaled into the dimensionless variable \(0 < \tilde{E} < \varepsilon_d/k_BT_{r0}\). In addition to the transfer processes between the reservoir and dimple states, the model also accounts for collisions between thermal atoms that remain within the thermal cloud. These collisions are assumed to occur on a timescale that is sufficiently short to maintain a quasiequilibrium distribution of the thermal component throughout the evolution.

Three-body recombination is a major loss mechanism in dense ultracold gases, particularly within the high-density region of the dimple trap. In this process, three atoms collide simultaneously, with two forming a bound molecule and the third carrying away the excess binding energy. The released kinetic energy typically exceeds the trap depth, causing all three atoms to leave the system. Three-body losses at sufficiently low temperatures can be accurately described as being proportional to the local probability of finding three atoms at the same position~\cite{pethick_smith_2008}. Following Ref.~\cite{PhysRevA.91.013601}, we write the three-body loss rate of the atomic density \(n(\mathbf r)\) as
\begin{equation}
\frac{dn(\mathbf{r})}{dt} \Bigg|_{\text{3-losses}} = -L \langle [\hat{\Psi}^\dagger(\mathbf{r})]^3 [\hat{\Psi}(\mathbf{r})]^3\rangle,
\end{equation}
where \(L\) is the three-body loss coefficient and \(\hat{\Psi}(\mathbf{r})\) is a sum of a condensate mean field \(\hat{\psi}_0(\mathbf{r})\) and a thermal fluctuations field \(\hat{\psi}_{th}(\mathbf{r})\). For \({}^{87}\mathrm{Rb}\), this coefficient has been measured experimentally to be \(L = 1.8 \times 10^{-29}\,\mathrm{cm}^6\,\mathrm{s}^{-1}\) \cite{Söding1999}.

The atomic density is decomposed into a condensate contribution \(n_0(\mathbf{r})= |\psi_0(\mathbf{r})|^2\) and an excited-state component \(n_{\mathrm{ex}}(\mathbf{r}) = \langle\hat{\psi}_{th}^\dagger(\mathbf{r})\hat{\psi}_{th}(\mathbf{r})\rangle\). Substituting this decomposition into the above expression and applying Wick’s theorem to evaluate the expectation value, the loss rate becomes
\begin{align}
\frac{dn(\mathbf{r})}{dt} \Bigg|_{\text{3-losses}}
= -L \Big[
& n_0^3(\mathbf{r})
+ 9\,n_0^2(\mathbf{r})\,n_{\mathrm{ex}}(\mathbf{r}) \nonumber \\
& + 18\,n_0(\mathbf{r})\,n_{\mathrm{ex}}^2(\mathbf{r})
+ 6\,n_{\mathrm{ex}}^3(\mathbf{r})
\Big].
\end{align}

\textcolor{black}{Our model assumes a fixed effective dimple volume (\(l_d^3\)) and omits mean-field expansion. In the Thomas-Fermi regime repulsive interactions broaden the condensate and lower the peak density relative to the fixed-volume assumption, which reduces three-body loss rates (\(\sim n^3\)) stronger than rates of elastic collisions (\(\sim n^2\)). This would favor BEC formation over atom loss and make BEC production dynamics even more efficient. A self-consistent description coupling spatial dynamics to kinetic thermalization is beyond the scope of this work; however, since the atom number in the BEC remains low, the present framework reliably captures the core thermalization kinetics and qualitative performance trends across the extraction protocols.}

\subsection{Repeated extraction and reservoir replenishment}
Extraction and replenishment are incorporated at the same coarse-grained level as the reservoir-dimple kinetic model. The dynamical variables are the condensate population, the energy-resolved noncondensed dimple distribution, and the thermodynamic variables of a quasi-equilibrated reservoir. Accordingly, atom extraction and external loading are represented by population and energy source terms acting on these variables, rather than by resolving the full spatio-temporal density and phase dynamics during a specific outcoupling or loading process.

This description is appropriate when the extraction or loading step is short compared with the subsequent collisional relaxation and condensate-growth timescales. In this regime, an extraction pulse primarily changes the dimple populations and prepares a new nonequilibrium initial state for the next growth stage. The subsequent evolution is then governed by the same reservoir-dimple kinetic processes as in the underlying model. Thus, the purpose of the present extension is not to optimize a particular experimental outcoupler, but to determine how repeated controlled depletion and replenishment affect the cycle-to-cycle production of condensed atoms.

Within this framework, atom extraction is described by the following phenomenological loss terms:
\begin{equation}
        \frac{\partial f_0}{\partial t} \Bigg|_{\text{extr}} = -\Gamma_0 f_0,
        \label{eqf0}
\end{equation}
\begin{equation}       
    \frac{\partial f(\tilde{E},t)}{\partial t} \Bigg|_{\text{extr}} =  - \Gamma D(\tilde{E})f(\tilde{E},t),
\end{equation}
\begin{equation}
    D(\tilde{E}) = \frac{2}{\sqrt{\pi}} \left( \frac{k_B T_{r0}m l_d^2}{ 2\pi\hbar^2}\right)^{3/2}\sqrt{\tilde{E}},
\end{equation}
where \(D(\tilde E)\) is the density of states for the chosen dimple model. The parameters \(\Gamma_0\) and \(\Gamma\) are the extraction-rate coefficients for the condensate and noncondensed dimple populations, respectively. Different extraction protocols are implemented by varying these two rates. In all simulations, the extraction pulse duration is fixed at \(0.1\,{\rm s}\).

In Sec.~\ref{bec_extr}, we compare three extraction protocols that differ in the relative depletion of the condensate and noncondensed dimple populations. In the Full-Clearance protocol, both \(\Gamma_0\) and \(\Gamma\) are chosen large enough that the dimple is effectively emptied during each extraction pulse. In the BEC-Clearance protocol, \(\Gamma_0\) remains large while \(\Gamma=0\), so that the condensate is removed but the noncondensed dimple population is retained. In the Partial-BEC-Clearance protocol, \(\Gamma_0\) is reduced and \(\Gamma=0\), leaving both a residual condensate and a noncondensed population in the dimple after the pulse. After each extraction pulse, collisional transfer from the reservoir repopulates the dimple, which allows repeated extraction cycles to be followed within a single simulation.

External reservoir replenishment is included by adding input contributions to the evolution equations for the reservoir population and energy. We assume that the number of atoms added to the reservoir via the external input is small compared to the total reservoir population \(N_{r0}\). Under this condition, the input contributions can be written as
\begin{equation}
    \frac{\partial f_r}{\partial t} \Bigg|_{\text{input}} =  \frac{1}{N_{r0}}\frac{\partial N_{\text{in}}(t)}{\partial t},
\end{equation}
\begin{equation}
     \frac{\partial e_r}{\partial t}\Bigg|_{\text{input}} = \frac{1}{E_{r0}}\frac{\partial E_{\text{in}}(t)}{\partial t},
\end{equation}
where \(N_{\text{in}}(t)\) and \(E_{\text{in}}(t)\) denote the number and total energy of the injected atoms, respectively. We further assume that the injected atoms enter the reservoir in the same thermal state as the initial cloud, implying that the injected energy per particle matches that of the initial reservoir. Consequently, the energy and particle inputs scale proportionally, and the energy input term can be written in the same form as the particle input,
\begin{equation}
     \frac{\partial e_r}{\partial t}\Bigg|_{\text{input}} = \frac{1}{N_{r0}}\frac{\partial N_{\text{in}}(t)}{\partial t}.
\end{equation}

\textcolor{black}{The extraction and replenishment processes introduced above can be implemented with weak perturbations of the trapped cloud, especially in repeated operation in which the different processes are implemented at different times and can be optimized independently. Reservoir replenishment can be performed along the elongated axis by locally lowering the trapping potential at the ends with large \(|z|\), where the density and associated losses are low; the small input rate considered below further limits the perturbation of the reservoir. Selective dimple extraction can be realized by fully or partially transferring atoms from the dimple to a movable optical tweezer of variable depth which is shuttled away from the dimple and reservoir location~\cite{Schlosser_2012,PhysRevResearch.5.L032009}. These mechanisms provide feasible physical realizations of the processes considered in the model.}

\subsection{BEC formation dynamics}
 
\begin{figure}[htb]
     \centering \includegraphics[width=\columnwidth]{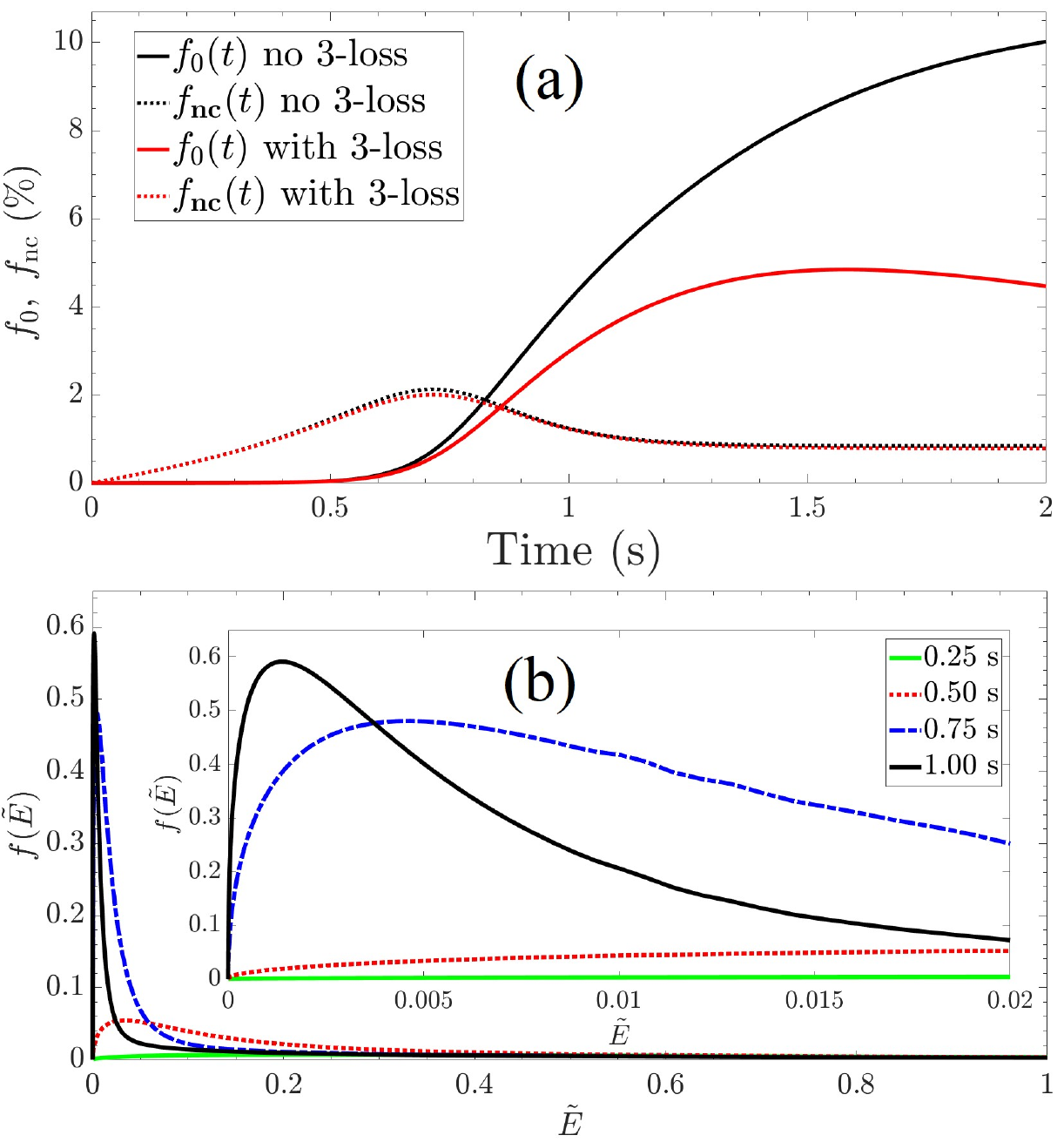}
\caption{
(a) Time evolution of the condensate fraction \(f_0(t)\) (solid) and the noncondensed dimple fraction \(f_{\rm nc}(t)\) (dotted). Red and black curves correspond to calculations with and without three-body losses. Evaporation is suppressed by taking \(\varepsilon_t\rightarrow\infty\). (b) Energy distribution \(f(\tilde E)\) of noncondensed atoms in the dimple at selected times for the calculation with three-body losses. The inset resolves the low-energy part of the distribution.
}\label{comp-3-loss}
 \end{figure}

For different initial conditions, the quantitative details of the dynamics vary, while the overall behavior remains qualitatively similar. The black curves in Fig.~\ref{comp-3-loss}(a) show the time evolution of the condensate and noncondensed fractions in the absence of three-body losses. The noncondensed fraction is defined as
\begin{equation}
f_{\rm nc} = \int_0^{\varepsilon_d/k_B T_{r0}} f(\tilde{E}, t) d\tilde{E}.
\end{equation}

After the dimple is switched on at \(t = 0\), atoms from the thermal cloud scatter into the dimple via elastic collisions. Once a sufficiently high phase-space density is reached, Bose-Einstein condensation occurs, accompanied by rapid growth of the condensate fraction and a corresponding decrease in the noncondensed population inside the dimple. The subsequent saturation of the condensate fraction takes place on a significantly longer timescale.
 
The dynamics are substantially modified in the presence of three-body losses, shown by the red curves in Fig.~\ref{comp-3-loss}(a). After Bose-enhanced growth accelerates condensate formation, the increasing density inside the dimple enhances the three-body recombination rate. As a result, the condensate acquires a finite lifetime and the condensate fraction develops a maximum before decaying. In contrast, the dynamics of the noncondensed fraction remain only weakly affected, with the largest deviations occurring near the onset of condensation.

We also analyze the time evolution of the atomic energy distribution inside the dimple. Representative snapshots obtained in the presence of three-body losses are shown in Fig.~\ref{comp-3-loss}(b). The figure presents the distribution function \(f(\tilde{E})\) at four different times. Here, \(\tilde{E}=0\) corresponds to the bottom of the dimple potential, while the maximum value \(\varepsilon_d/k_B T_{r0}\) corresponds to the matching point with the thermal reservoir. Shortly after the dimple is activated, most dimple atoms occupy low-energy states. The resulting sharply peaked Bose–Einstein distribution reflects the presence of a condensate in the system. Consequently, during most of the evolution the average energy of atoms inside the dimple remains close to the ground-state energy. This observation implies that excessively deep dimples are unfavorable for repeated condensate extraction, since they lead to significant heating of the thermal reservoir.

Finally, we examine the role of evaporation in the system dynamics. Although this mechanism is related to the standard evaporative cooling process commonly used to produce Bose-Einstein condensates, it does not constitute the dominant condensation mechanism in the present system. Because the reservoir trap depth is finite, sufficiently energetic thermal atoms can escape the trap during collisions. The preferential loss of high-energy atoms leads to cooling of the thermal cloud, and even relatively weak cooling can noticeably enhance condensate formation. The evaporation rate is determined by the maximum allowed energy of atoms in the reservoir, characterized in our model by \(\varepsilon_t\). Since the thermal cloud is assumed to remain in quasiequilibrium throughout the evolution, varying \(\varepsilon_t\) during the simulation would artificially modify the cloud temperature. For this reason, \(\varepsilon_t\) is kept constant in all simulations.

\section{Repeated BEC extraction}
\label{bec_extr}

\subsection{Repeated extraction without replenishment}

\begin{figure*}[htb] 
    \centering
    \includegraphics[width=0.9\textwidth]{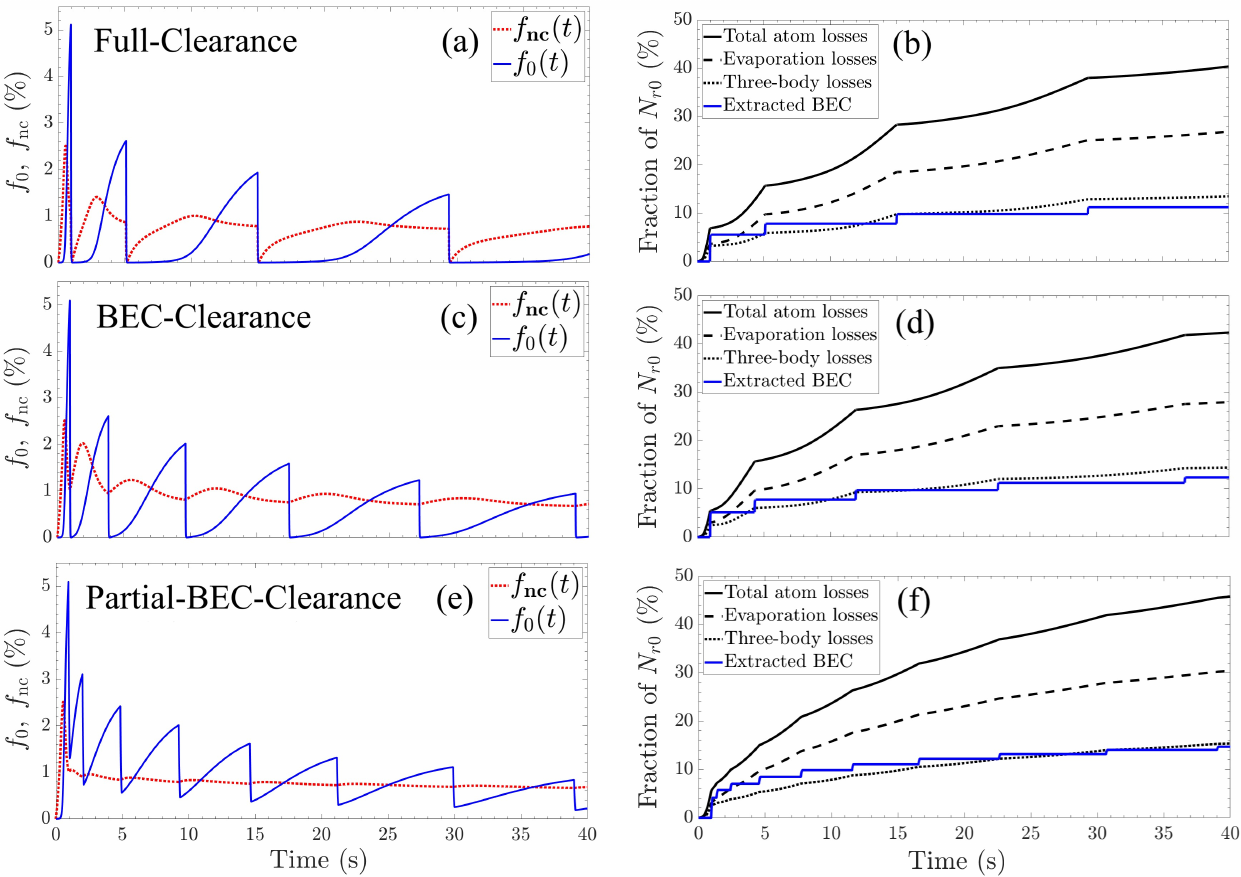}
    \caption{ 
    Repeated condensate extraction without external replenishment. Left panels show the condensate fraction \(f_0(t)\) and noncondensed dimple fraction \(f_{\textbf{nc}}(t)\). Right panels show the cumulative particle balance. The reservoir depth is \(\varepsilon_t=k_B\times750\,{\rm nK}\). The critical temperature in the absence of the dimple is \(T_c=130\,{\rm nK}\). (a),(b) Full-Clearance protocol. (c),(d) BEC-Clearance protocol. (e),(f) Partial-BEC-Clearance protocol. 
}
    \label{protocol-comp}
\end{figure*}

We next investigate repeated condensate extraction from the dimple without external replenishment of the thermal reservoir. Since no atoms are injected into the system, each extraction irreversibly depletes the reservoir and modifies its temperature. Extraction pulses are applied when the condensate fraction reaches  \SI{90}{\percent} of the maximum value obtained in the absence of extraction, so that all protocols share the same first extraction time. Figure~\ref{protocol-comp} summarizes the resulting dynamics and cumulative particle balance.

We compare three extraction protocols that differ in the amount of the dimple population retained after each pulse. In the Full-Clearance protocol, both condensate and noncondensed dimple populations are removed after extraction. This protocol is implemented by setting \(\Gamma_0=100\,{\rm s}^{-1}\) and \(\Gamma=300\,{\rm s}^{-1}\). As a result, the dimple must be repopulated from the reservoir before Bose-stimulated growth can resume. This leads to progressively longer delays between successive extractions and limits the total extracted condensate fraction to \SI{10.8}{\percent} of \(N_{r0}\).

In the BEC-Clearance protocol, only condensate atoms are extracted, while the noncondensed dimple population remains in the system. This protocol is implemented by setting \(\Gamma_0=100\,{\rm s}^{-1}\) and \(\Gamma=0\). The residual thermal component substantially accelerates subsequent condensate formation by reducing the delay before Bose-enhanced growth begins anew. Consequently, additional extractions become possible before reservoir depletion suppresses condensation, increasing the total extracted condensate fraction to \SI{13.6}{\percent} of \(N_{r0}\). Simultaneously, the persistent dimple population enhances evaporation and three-body recombination losses.

The shortest recovery times are obtained in the Partial-BEC-Clearance protocol, where both noncondensed atoms and part of the condensate remain in the dimple after each extraction. This protocol is implemented by setting \(\Gamma_0=15\,{\rm s}^{-1}\) and \(\Gamma=0\). In this regime, the residual condensate immediately seeds further Bose-stimulated growth, allowing extractions to occur at significantly shorter intervals. Although the increased dimple density strengthens three-body losses, the cumulative extracted condensate fraction reaches \SI{13.9}{\percent} of \(N_{r0}\), which is the highest value among the three protocols.

\begin{figure}[t]
    \centering \includegraphics[width=\columnwidth]{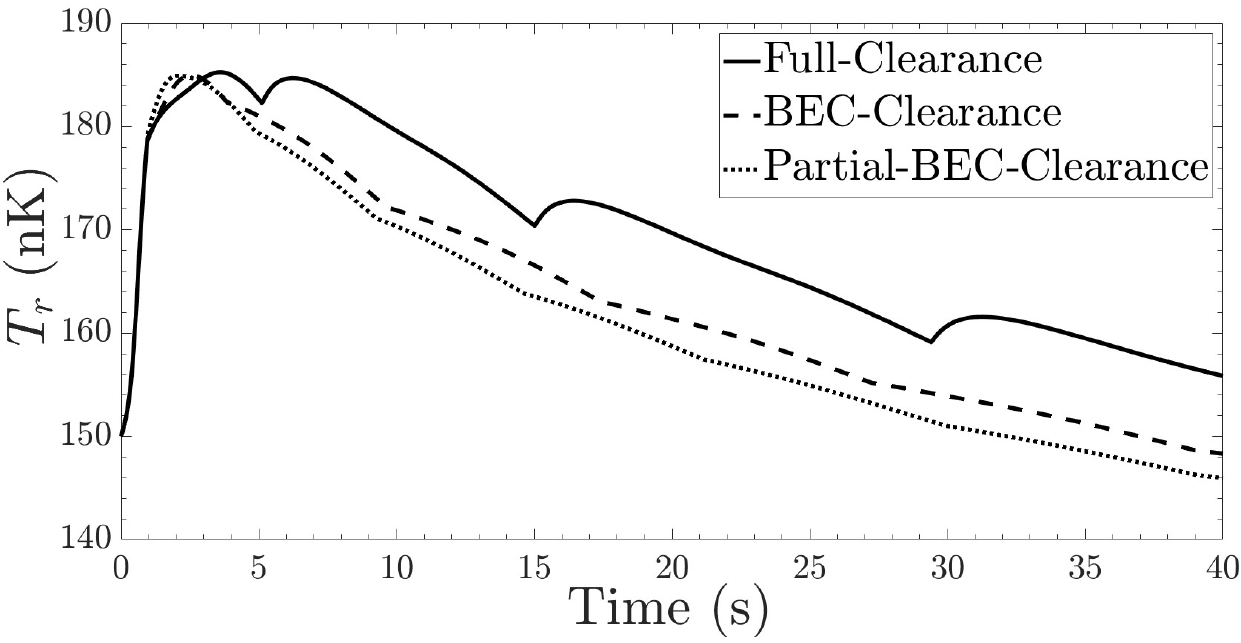}
\caption{
Reservoir temperature as a function of time for the extraction protocols shown in Fig.~\ref{protocol-comp}. Condensate formation initially heats the reservoir, whereas subsequent evaporation from the finite-depth trap gradually lowers the temperature.
}   \label{T-vs-protocol}
\end{figure}

For long-term dynamics, controlling the thermal cloud temperature is essential and requires sufficiently strong evaporation to offset the heating induced by repeated condensation cycles. Since the dimple lies below the thermal cloud on the energy scale, condensate growth heats the thermal reservoir while reducing its atom number. Consequently, subsequent BEC extractions occur at higher temperatures and with fewer thermal atoms, reducing the peak condensate fraction. Three-body losses further enhance both heating and atom depletion. To sustain continued condensation, evaporation must therefore be enhanced by lowering the reservoir trap depth. In our case, the trap  depth \(\varepsilon_t = k_B \times 750\text{ nK}\) provides sufficient cooling, as shown in Fig.~\ref{T-vs-protocol}. The temperature initially increases by about 30 nK within the first second (i.e. the first cycle of BEC formation) and then gradually returns toward its initial value due to enhanced evaporation. The BEC-Clearance protocol maintains a lower average temperature (reduced by about 8 nK relative to the Full-Clearance protocol) because the persistent noncondensed fraction in the dimple increases evaporation. The Partial-BEC-Clearance protocol further reduces the temperature by approximately 2 nK due to the additional presence of condensate atoms. For substantially larger trap depths, the resulting heating would suppress condensation after only the first few extractions.

\begin{figure}[htb]
    \centering \includegraphics[width=\columnwidth]{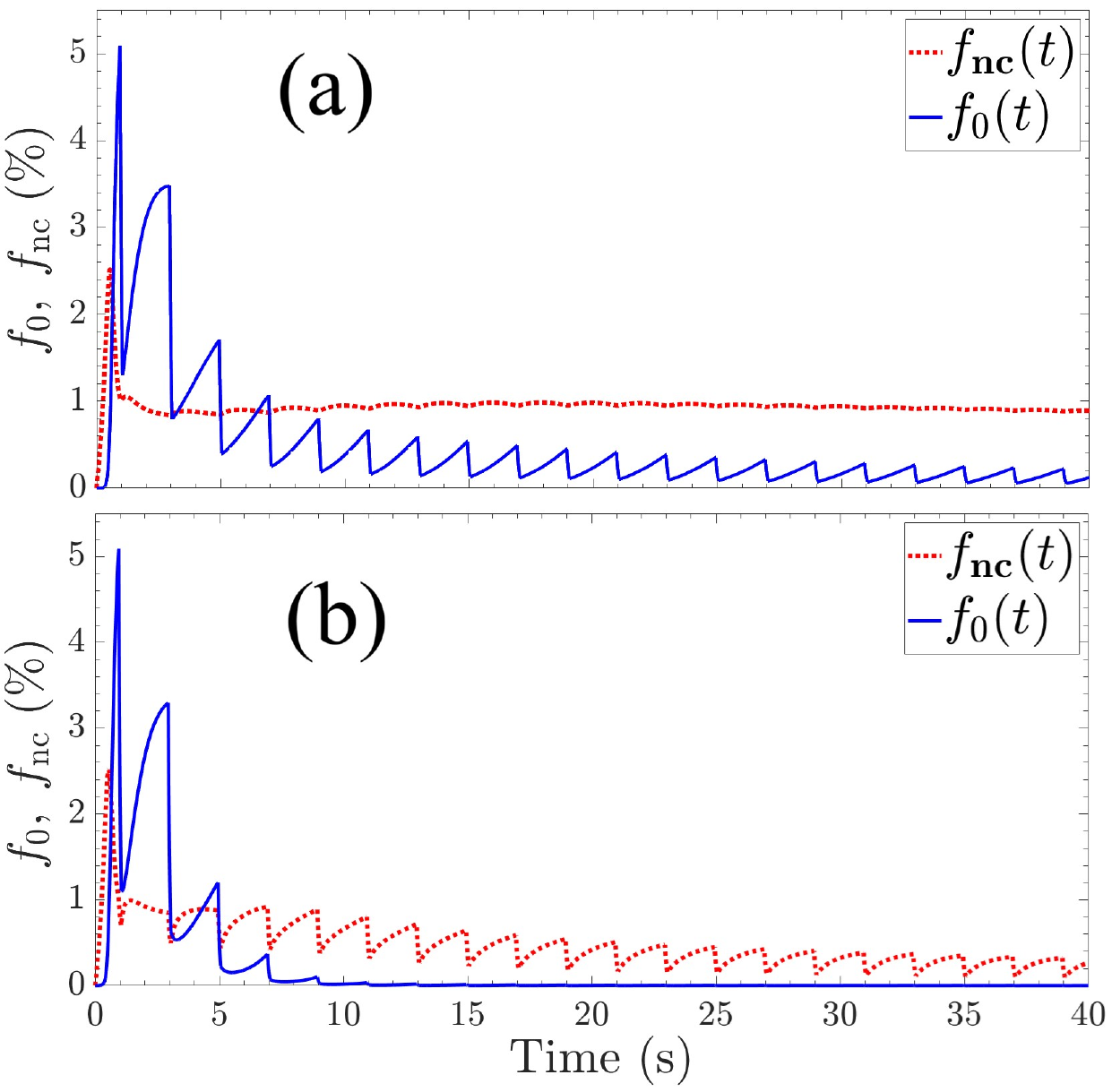}
\caption{Fixed-period extraction without external replenishment. After the first extraction, subsequent extraction pulses are applied with period \(T_{\rm extr}=2\,{\rm s}\). Solid blue curves denote the condensate fraction \(f_0(t)\), and dotted red curves denote the noncondensed dimple fraction \(f_{\textbf{nc}}(t)\). (a) Partial-BEC-Clearance protocol with \(\Gamma_0=15\,{\rm s}^{-1}\) and \(\Gamma=0\) . (b) Modified protocol including partial noncondensed-atom clearance with \(\Gamma_0=15\,{\rm s}^{-1}\) and \(\Gamma=20\,{\rm s}^{-1}\).
}
\label{plot4}
\end{figure}

The above analysis focuses on the long-term dynamical regime. However, in systems with a limited condensate lifetime, multiple extractions may be impractical. In this case, it is more relevant to focus on the earliest condensations, which require the least amount of time. For the Full-Clearance protocol, only two extractions can be performed efficiently, extracting \SI{7.5}{\percent} of \(N_{r0}\) as condensate atoms in 5.2 s (including the extraction time). For the BEC-Clearance protocol, two extractions can be performed efficiently as well, extracting \SI{7.8}{\percent} of \(N_{r0}\) as condensate atoms in 3.95 s. On these timescales, the reduction in recovery time becomes particularly significant. For the Partial-BEC-Clearance protocol, three extractions can be performed, extracting \SI{7.0}{\percent} of \(N_{r0}\) as condensate atoms in 4.95 s. Thus, at the expense of a reduction in cumulative extracted BEC, a larger number of short-term experiments can be realized.

A clear disadvantage of schemes without external atom replenishment is the unpredictability of the next extraction time. This limitation can be reduced by using a variant of the Partial-BEC-Clearance protocol, in which residual condensate and noncondensed atoms remain simultaneously in the dimple, allowing extraction to be performed over a broader temporal interval. This introduces a degree of experimental control, enabling extraction at regular intervals after the first event, as illustrated in Fig.~\ref{plot4}(a). Such fixed-period extraction, together with the persistent thermal background in the dimple, can establish a quasi-periodic small-amplitude regime. However, this regime is sustained only if most noncondensed atoms remain in the dimple; significant depletion of this component quickly suppresses the effect, as shown in Fig.~\ref{plot4}(b).

The above results lead to several conclusions. The Full-Clearance and BEC-Clearance protocols define lower and upper limits for all schemes in which the condensate is completely extracted from the system. Since the difference between them is noticeable but not dramatic, it can be concluded that restricting extraction to the condensate alone does not significantly improve the overall performance; rather, it is mainly relevant for the purity of the extracted BEC. Moreover, the Partial-BEC-Clearance protocol, which in the analysis appears as a potential improvement over the BEC-Clearance protocol, shows that avoiding complete removal of the condensate can improve long-term performance. However, in the short-term regime, due to increased losses associated with higher density in the dimple, the Partial-BEC-Clearance protocol is somewhat less efficient than the BEC-Clearance protocol. This situation changes in the presence of external atomic replenishment.

\subsection{Fixed-period extraction with continuous replenishment}
\label{ext_input}
\begin{figure*}[t] 
    \centering
\includegraphics[width=0.9\textwidth]{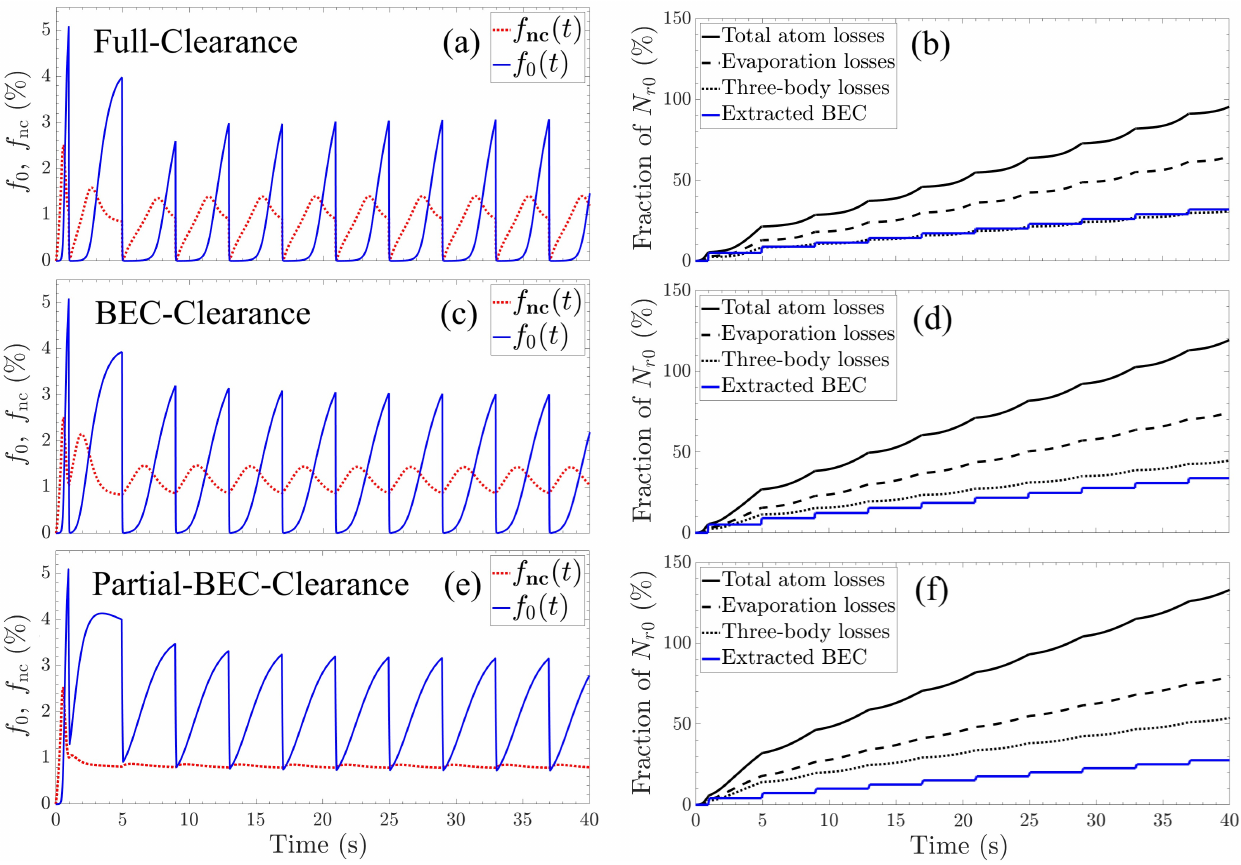}
 \caption{
 Fixed-period condensate extraction with continuous reservoir replenishment for \(\dot N_{\rm in}=10^{4}\,{\rm s}^{-1}\), \(T_{\rm extr}=4.0\,{\rm s}\), and \(\varepsilon_t=k_B\times750\,{\rm nK}\). Left panels show the condensate fraction \(f_0(t)\) and noncondensed dimple fraction \(f_{\textbf{nc}}(t)\). Right panels show the cumulative particle balance. (a),(b) Full-Clearance protocol. (c),(d) BEC-Clearance protocol. (e),(f) Partial-BEC-Clearance protocol.
}
\label{external-comp}
\end{figure*}

\begin{figure*}[t] 
    \centering
\includegraphics[width=0.95\textwidth]{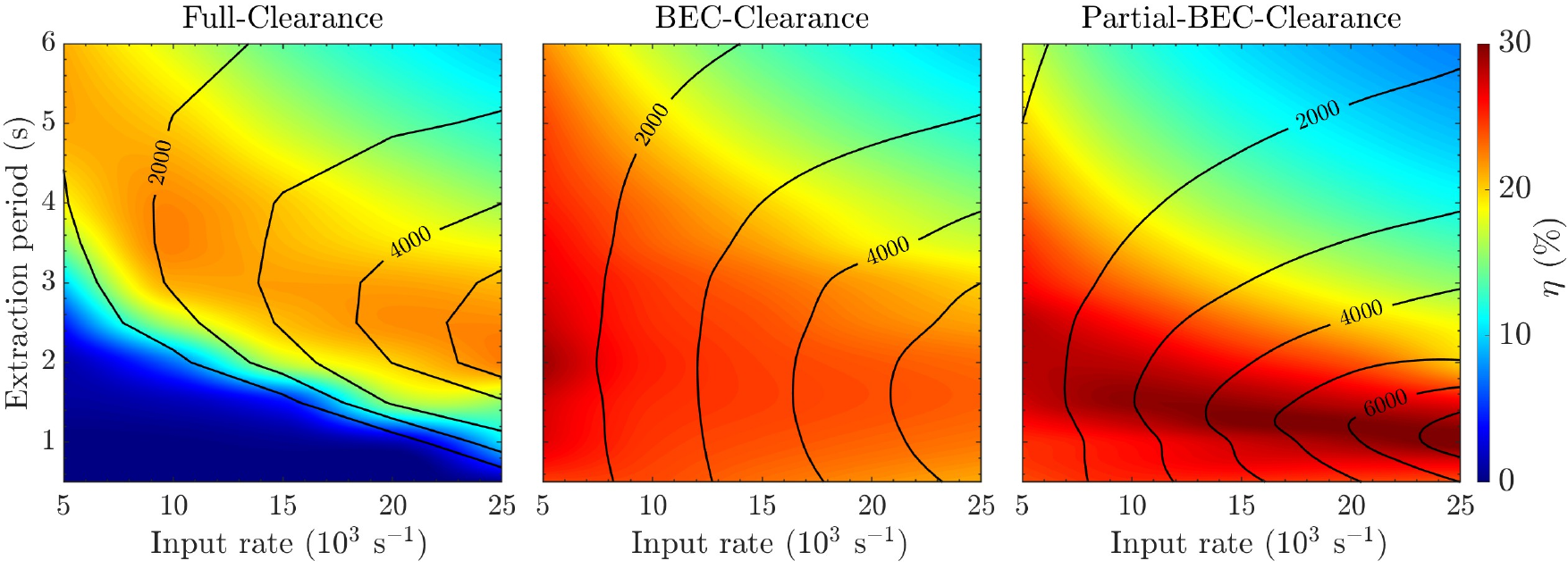}
\caption{
\textcolor{black}{Replenishment efficiency \(\eta\) (color-coded density plot) and average extracted-condensate output rate \(\eta\dot N_{\rm in}\) (black contour lines) for the three extraction protocols as a function of the thermal-atom input rate \(\dot N_{\rm in}\) and extraction period \(T_{\rm extr}\).
}
}
    \label{eff_maps}
\end{figure*}

We now consider the behavior of the three protocols with fixed-period extraction when atoms are continuously injected into the reservoir. As shown in Fig.~\ref{protocol-comp}, without replenishment of thermal atoms, successive extractions occur at different reservoir atom numbers and temperatures, making both the onset time of the next condensation cycle and the extracted condensate fraction difficult to predict. Introducing an external source of thermal atoms compensates for atom losses and can restore consistent cycle-to-cycle dynamics.

Figure \ref{external-comp} summarizes the repeated operation of the extraction protocols in the presence of a continuous input of thermal atoms into the reservoir. The model parameters and extraction protocols are the same as in the preceding subsection, except that continuous reservoir replenishment is now included. Specifically, the thermal-atom input rate is \(\dot N_{\rm in}=10^{4}\,{\rm s}^{-1}\), switched on at \(t=0.95\,{\rm s}\), and the extraction period is fixed at \(T_{\rm extr}=4.0\,{\rm s}\) for all protocols. As shown in Figs.~\ref{external-comp}(a), (c), and (e), after a short transient regime all cases approach a cyclic regime: the condensate is periodically depleted by extraction and subsequently restored through loading and rethermalization, while the noncondensed component exhibits a phase-shifted response. The use of external replenishment significantly changes the behavior of subsequent extractions. The extractions become reproducible and, importantly, require only a few cycles to come close to the asymptotic regime. Moreover, this is achieved with an input rate that, over one cycle, adds a number of atoms of \(4\times10^{4}\), which remains small compared with the initial thermal reservoir population \(N_{r0}=3\times10^5\). Therefore, if atomic replenishment can be implemented experimentally, it would significantly simplify the realization of repeated extractions.

The resulting dynamics differ between protocols. In the Full-Clearance protocol, the noncondensed fraction accumulates during each cycle until it reaches a value of slightly above \SI{1}{\percent}, after which Bose-stimulated growth leads to condensate formation. Importantly, the extraction period is sufficiently long for condensation to occur. If the period were shorter than the saturation time of the dimple with noncondensed atoms, the extraction efficiency would approach zero. For the BEC-Clearance protocol, such a characteristic time also exists; however, unlike in the previous case, it is substantially reduced. The noncondensed fraction must still reach saturation, but because the dimple already contains a significant number of atoms, this occurs more rapidly. In contrast, for the Partial-BEC-Clearance protocol, no such waiting time exists. In this case, condensate formation begins immediately at a high rate.

The efficiencies of the asymptotic periodic cycles differ between protocols. The Full-Clearance protocol converts \SI{22.3}{\percent}  of the injected thermal atoms into extracted condensate atoms, while the BEC-Clearance and Partial-BEC-Clearance regimes reach \SI{22.8}{\percent} and \SI{18.5}{\percent}, respectively. In comparison, the case without external atom replenishment yielded no more than 14 \% of the initial cloud extracted as condensate, demonstrating that periodic extraction combined with continuous replenishment substantially improves the long-term performance of the system. This comparison was carried out for identical extraction periods and thermal atom input rates. Under these conditions, the BEC-Clearance and especially the Partial-BEC-Clearance protocols are more strongly affected by additional losses associated with incomplete removal of the thermal component, as seen in Fig. \ref{external-comp}(b), (d), and (f). Although the reduction in efficiency remains relatively small in the BEC-Clearance regime, it becomes substantial for the Partial-BEC-Clearance protocol. This motivates a systematic comparison across the parameter space defined by the input rate and the extraction period.
 
For further analysis, we introduce the replenishment efficiency \(\eta\), defined as the average number of condensate atoms extracted from the system divided by the number of injected thermal atoms:
\begin{equation}
     \eta = \frac{N_{\text{extr}}^{\text{BEC}}(t_0+T_{\text{extr}}) - N_{\text{extr}}^{\text{BEC}}(t_0) }{N_{\text{in}}(t_0+T_{\text{extr}}) - N_{\text{in}}(t_0)} ,
\end{equation}
where \(t_0\) is the beginning of the averaging interval after the system reaches a quasi-stationary cyclic regime, \(T_{\text{extr}}\) is the extraction period, and \(N_{\text{extr}}^{\text{BEC}}(t)\) and \(N_{\text{in}}(t)\) are the total number of extracted condensate atoms and injected thermal atoms, respectively. Thus, the parameter \(\eta\) quantifies the efficiency of converting the injected thermal input atoms into extracted condensate atoms. 

We constructed heatmaps of the replenishment efficiency \(\eta\) for each extraction protocol, as shown in Fig.~\ref{eff_maps}. The efficiency was evaluated over the parameter space defined by the thermal-atom replenishment rate and the extraction period. The parameter scans covered input rates \(5\times10^{3}\,{\rm s}^{-1}\leq \dot N_{\rm in}\leq 2.5\times10^{4}\,{\rm s}^{-1}\), in steps of \(5\times10^{3}\,{\rm s}^{-1}\), and extraction periods \(0.5\,{\rm s}\leq T_{\rm extr}\leq 6.0\,{\rm s}\), in steps of \(0.5\,{\rm s}\). For the Partial-BEC-Clearance protocol, an additional point at \(T_{\rm extr}=1.25\,{\rm s}\) was included to improve the resolution near the efficiency maximum. The selected parameter range captures the relevant dynamical regimes of the system, while regions corresponding to very short extraction periods and very low replenishment rates were excluded because they produced cycles involving fewer than 1000 transferred atoms and therefore had limited physical relevance. For each parameter set, the dynamics were evolved until the cycle-averaged replenishment efficiency satisfied \({|\eta(t_0+T_{\rm extr})-\eta(t_0)|}/{\eta(t_0)}<10^{-3}\). The resulting data were interpolated onto a \(100\times100\) grid to generate the heatmaps, while black contour lines indicate the average output rate of condensate atoms. 

A characteristic feature of all heatmaps is the presence of a single local maximum both along lines of constant extraction period and along lines of constant replenishment rate. This behavior implies that, for a fixed replenishment rate, an optimal extraction period exists that maximizes the replenishment efficiency, while for a fixed extraction period there is also an optimal thermal input rate. The decrease in efficiency observed at sufficiently long extraction periods and at excessively high replenishment rates originates from enhanced three-body losses. For long extraction periods, the additional waiting time initially promotes condensate growth, but beyond a certain point the increase in condensate population is outweighed by three-body recombination losses, reducing the overall efficiency. Similarly, increasing the replenishment rate initially enhances condensate formation by supplying additional atoms, but the resulting increase in density amplifies three-body losses. Above a certain input rate, the growth of losses exceeds the gain in extracted condensate atoms, leading to a reduction in efficiency. These results demonstrate that each extraction protocol possesses a distinct region of optimal operation in the parameter space of extraction period and external replenishment rate.

The Full-Clearance protocol exhibits low replenishment efficiency for \(T_{\rm extr}<2.0\,{\rm s}\), and no efficient operation is obtained for \(T_{\rm extr}<1.0\,{\rm s}\). This behavior reflects the finite waiting time required to repopulate the dimple and initiate Bose-stimulated growth after each complete clearance. For longer extraction periods, the protocol shows a broad region with \(\eta>\SI{20}{\percent}\), but the maximum efficiency remains limited to approximately \SI{22.5}{\percent}.  

The BEC-Clearance protocol reduces this waiting time by retaining the noncondensed dimple population after each pulse. As a result, efficient operation extends to shorter extraction cycles than in the Full-Clearance case. The peak efficiency increases substantially, reaching approximately \SI{29}{\percent}. However, the region of parameter space where such high efficiencies are achieved remains relatively narrow. For the Partial-BEC-Clearance protocol, the optimal region is shifted toward shorter extraction periods. The efficiency decreases for \(T_{\rm extr}\gtrsim 3.0\,{\rm s}\), where the longer residence time in the dimple enhances density-dependent losses. By contrast, short periods around \(T_{\rm extr}\simeq 1.0\,{\rm s}\) give a noticeable increase in efficiency. Although the maximum efficiency is only slightly higher than in the BEC-Clearance protocol, the Partial-BEC-Clearance protocol exhibits a substantially broader high-efficiency region, reaching a maximum replenishment efficiency of approximately \SI{30}{\percent}.

The three extraction protocols studied here differ significantly in their dynamical behavior. The Full-Clearance protocol performs best when a balance is maintained between the replenishment rate and the extraction period, and it is particularly sensitive to short extraction periods, which suppress condensate formation. This limitation is not present in the BEC-Clearance protocol due to the continuous presence of a noncondensed component, which extends efficient operation to shorter extraction periods and increases the maximum achievable efficiency. The Partial-BEC-Clearance protocol exhibits the highest replenishment efficiency among all protocols, with its maximum located in the short-period region.
\textcolor{black}{This improvement is a direct consequence of residual-BEC-assisted growth: the remaining condensate eliminates the waiting time for spontaneous dimple replenishment, albeit at the cost of enhanced density-dependent losses. Consequently, Partial-BEC-Clearance performs best for thermal-atom replenishment at high input rates and short extraction periods, which is advantageous for systems with limited BEC lifetime. }

\subsection{Comparison with direct-current (DC) operation}

\begin{figure}[t]
     \centering \includegraphics[width=\columnwidth]{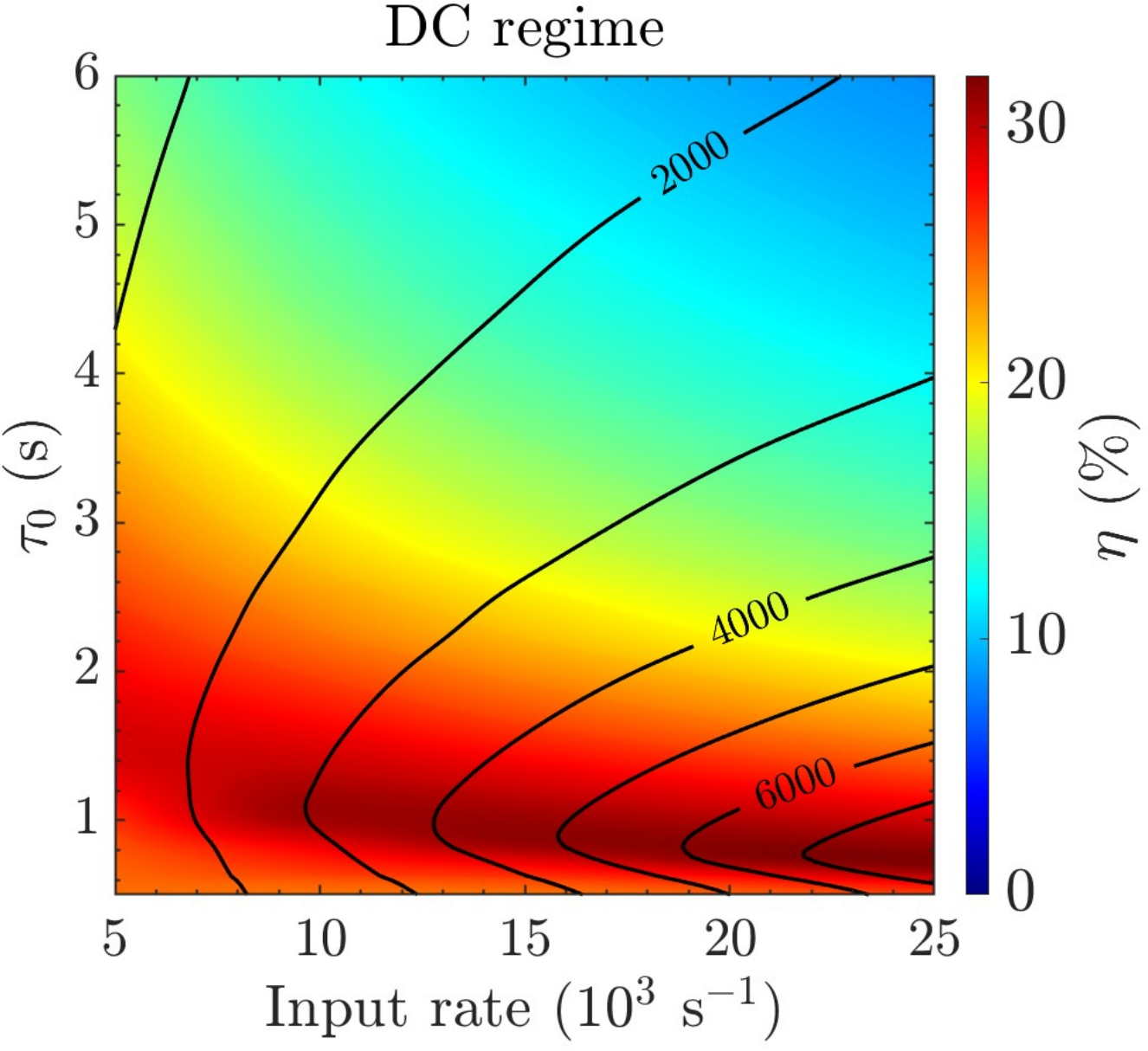}
\caption{\textcolor{black}{Replenishment efficiency \(\eta\) (density plot) and average extracted-condensate output rate \(\eta\dot N_{\rm in}\) (contour lines) for direct-current (DC) operation as a function of the thermal-atom input rate \(\dot N_{\rm in}\) and the condensate outcoupling timescale \(\tau_0 = \Gamma_0^{-1}\) given as the inverse of the extraction-rate coefficient for the condensate.}}
\label{dc}
\end{figure}

\textcolor{black}{
Although repeated BEC formation is the central topic of this article, we also examine a continuously acting extraction process for comparison. Therefore, we examine the direct-current (DC) regime in which the extraction terms continuously remain active. In this case, the condensate outcoupling timescale, rather than the extraction period, serves as the control parameter determining the steady-state condensate flux and replenishment efficiency.
} 

\textcolor{black}{
We constructed a heatmap of the replenishment efficiency \(\eta\) for the DC regime, analogous to those obtained for the periodic protocols, with the outcoupling timescale varied in place of the extraction period \(T_{\mathrm{extr}}\). Extraction is applied to the condensate only, without extraction of noncondensed atoms. The resulting efficiency map is shown in Fig.~\ref{dc} as a function of the thermal-atom replenishment rate \(\dot N_{\rm in}\) and the condensate outcoupling timescale \(\tau_0\). We define \(\tau_0=\Gamma_0^{-1}\), where \(\Gamma_0\) is the extraction-rate coefficient for the condensate introduced in Eq.~\eqref{eqf0}, following the conventional description of atom-laser output coupling~\cite{Hagley1999AtomLaser,Helmerson_1999,bolpasi:2014:atomlaser}. The parameter scans cover input rates \(5\times10^{3}\,{\rm s}^{-1}\leq\dot N_{\rm in}\leq2.5\times10^{4}\,{\rm s}^{-1}\), in steps of \(5\times10^{3}\,{\rm s}^{-1}\), and  outcoupling timescales \(0.5\,{\rm s}\leq\tau_0\leq6.0\,{\rm s}\), in steps of \(0.25\,{\rm s}\). For each parameter set, the dynamics were evolved until the relative change in the replenishment efficiency over a one-second interval was below \(10^{-3}\). The resulting data were interpolated onto a \(100\times100\) grid to generate the heatmap, while black contour lines indicate the average condensate output rate.}

\textcolor{black}{Within the explored parameter range, the replenishment efficiency and BEC output rate in the DC regime are comparable to those obtained for pulsed Partial-BEC-Clearance operation, with a maximum replenishment efficiency of approximately \SI{32}{\percent}. At their respective optima, the two modes of operation also exhibit a similar balance between condensate extraction and the dominant loss channels, suggesting that they approach a comparable upper efficiency limit under the conditions considered here. We note that this analysis compares separately optimized pulsed and DC operating regimes rather than demonstrating convergence of the pulsed protocol to the DC limit. This comparison shows that the introduced pulsed Partial-BEC-Clearance protocol is almost as efficient as operation in the DC regime, while offering the advantage that the different stages of the protocol, such as BEC formation and BEC extraction, can be separated temporarily and optimized independently. This presents a major advantage for implementing repeated and quasi-continuous schemes of BEC production. 
}

\section{Conclusions}\label{sec:conlc}

In this work, we investigated repeated Bose-Einstein condensate extraction in a reservoir-dimple system using a population-kinetic framework that includes Bose-stimulated growth, evaporation, three-body losses, and external reservoir replenishment. Three extraction protocols were compared, differing in the amount of the dimple population retained after each extraction.

The key physical point is that repeated extraction is not simply a sequence of independent condensate-growth events. Because residual dimple populations survive between cycles, \textcolor{black}{the remaining noncondensed and condensed fractions seed Bose-stimulated growth in the subsequent cycle}, shifting the optimal operating regime from isolated condensation toward a loss-limited cyclic steady state.

The results demonstrate that residual dimple populations play a central role in determining the efficiency of repeated condensate production. Retaining noncondensed atoms in the dimple substantially reduces the recovery time between extractions by accelerating the onset of Bose-stimulated growth, while leaving part of the condensate further enhances subsequent regrowth. These mechanisms increase the achievable condensate throughput and enable efficient operation at shorter extraction periods. At the same time, the persistent dimple population increases density-dependent losses through evaporation and three-body recombination, so the optimal operating regime is determined by the competition between accelerated recovery and enhanced losses.

\textcolor{black}{For continuous external replenishment, the Partial-BEC-Clearance protocol provides the highest replenishment efficiency among the pulsed extraction schemes considered here. Its advantage originates from the residual condensate, which immediately seeds Bose-stimulated growth after each extraction and thereby enables efficient operation at shorter extraction periods.}

\textcolor{black}{The present model describes the cycle-to-cycle population dynamics of the replenishment-extraction process, but does not include phase coherence, collective excitations, or the microscopic dynamics of an extracted BEC.}

\begin{acknowledgments}
K.K. acknowledges support from NFDU project No 0124U002616. M.E.\,is supported by U.S.\,National Science Foundation Grant No.\,PHY-2207476. We thank Y.M.\,Bidasyuk for helpful comments and suggestions that improved the clarity of the manuscript.
\end{acknowledgments}

\bibliographystyle{apsrev4-2}
\bibliography{ref}

\end{document}